# Distribution of Time-Energy Entanglement over 100 km fiber using superconducting single-photon detectors

Qiang Zhang[1], Hiroki Takesue[2], Sae Woo Nam[3], Carsten Langrock[1], Xiuping Xie[1], M. M. Fejer[1], Yoshihisa Yamamoto[1,4]

[1] *Edward L. Ginzton Laboratory, Stanford University, Stanford, California 94305*
[2] *NTT Basic Research Laboratories, NTT Corporation, 3-1 Morinosato Wakamiya, Atsugi, Kanagawa 243-0198, Japan*
[3] *National Institute of Standards and Technology, 325 Broadway, Boulder, Colorado 80305*
[4] *National Institute of Informatics, 2-1-2 Hitotsubashi, Chiyoda-ku, Tokyo, 101-843, Japan*
*qiangzh@stanford.edu*

**Abstract:** In this letter, we report an experimental realization of distributing entangled photon pairs over 100 km of dispersion-shifted fiber. In the experiment, we used a periodically poled lithium niobate waveguide to generate the time-energy entanglement and superconducting single-photon detectors to detect the photon pairs after 100 km. We also demonstrate that the distributed photon pairs can still be useful for quantum key distribution and other quantum communication tasks.

___

___________________________________________________________________________

**1. Introduction**

Entanglement distribution is one of the core components in the field of long-distance quantum communication (LDQC), for example, quantum key distribution (QKD) [1,2] and quantum teleportation [3,4]. So far there are many experimental implementations of entanglement distribution over free-space links [5,6], or optical fibers [7-11].

Very recently, 100-km entanglement distribution over optical fiber has been reported by several groups [9-11]. One of the main problems in long-distance entanglement distribution is the dark count rate of the detectors. Here, we utilized a reverse-proton-exchange (RPE) periodically poled lithium niobate (PPLN) waveguide and superconducting single-photon detectors (SSPD) to improve the distribution distance to 100 km over dispersion-shifted fiber (DSF). RPE PPLN waveguides [12] were used due to their high nonlinear efficiencies and low propagation loss (<0.1 dB/cm), while the SSPDs [13,14] had very low dark count rates (<200 Hz) and a fast timing response (~65 ps full width at half maximum), which resulted in a low dark count probability per time window. We demonstrated that the entanglement after 100 km still showed quantum nonlocality and violated Bell's inequality by observing the two-photon-interference fringes.

**2. Time-Energy Entanglement**

Time-energy-entangled photon pairs at telecom wavelengths are thought to be good candidates for LDQC over fiber-based networks [7-9] due to the low propagation loss in standard optical fiber and insensitivity to polarization-mode dispersion compared with polarization-entangled photon pairs. Time-energy entanglement was first proposed by J. D. Franson in 1989 [15]. In his original paper, an atom cascade process is utilized to generate the entanglement. However, current experimental realizations of the protocol are mainly based on parametric down conversion (PDC) [7].

As shown in Fig. 1, a continuous-wave laser with an ultra-long coherence time $\tau_1$ pumps a crystal possessing a $\chi^{(2)}$ nonlinearity, generating correlated photon pairs via the PDC process. The down-converted photon pairs usually have a much broader spectrum than the pump photons and hence a significantly shorter coherence time $\tau_2$ according to the time-bandwidth limit. Since the two photons of a photon pair are always generated simultaneously and the down conversion process obeys energy conservation, both the total energy and the time difference of the two photons are well defined as the pump photon's energy and zero, respectively. However, we do not know the exact time and energy of each photon. This kind of entangled state is called time-energy entanglement, and can be mathematically expressed as, $|\psi\rangle = C\int dt_1 dt_2 e^{-(t_1-t_2)^2/4\tau_2^2} e^{-(t_1+t_2)^2/16\tau_1^2} e^{-i(\omega_p/2)(t_1+t_2)} a_1^+(t_1)a_2^+(t_2)|0\rangle$, where C is a normalization constant, $w_p$ is the pumping frequency and $a_i^+(t_i)$ is the photon creation operator in mode i at time $t_i$ [16]. It has been widely used for testing of quantum nonlocality and quantum key distribution.

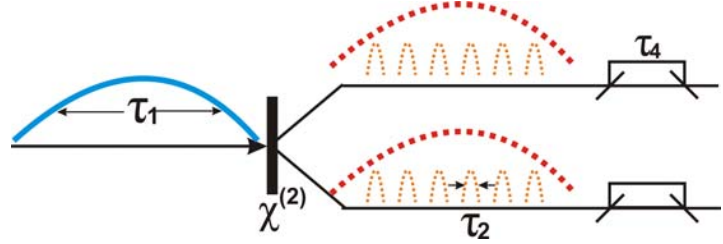

Fig. 1. Scheme of generation and detection of time-energy entanglement. The small dotted pulses represent the possible photon pairs, which can be generated during the long time span of the pump light, while the dotted red envelope of the photon pairs is from the long pump.

To characterize and utilize the entanglement, one usually launches the entangled photon pairs into two unbalanced Mach-Zehnder interferometers (MZI) and then implements a coincidence measurement by detecting the outputs of the two MZIs with single-photon detectors. Suppose the timing jitter of the single-photon detector is $\tau_3$ and the time unbalance between the MZI's two paths is $\tau_4$. When a photon with quantum state $|t\rangle$ passes through the MZI at time n, it will split, propagate in the short and long arms and come out with the quantum state $(|t\rangle + e^{i\theta}|t+\tau_4\rangle)/\sqrt{2}$, where $\theta$ is the phase between the two different arms. Suppose $\tau_1 \gg \tau_4 > \tau_2, \tau_3$, such that there will be no single-photon coherence fringe in the MZI's output. When a photon pair is detected in the two MZI's outputs at time $t + \tau_4$, it could either have been generated at time $t$, propagated in the MZI's long arm or could have been generated at a later time $t + \tau_4$, propagated in the short arm. Since the photon pairs generated during the pump light coherence time $\tau_1$ were all in phase and $\tau_1 \gg \tau_4$, the two possibilities were interfered and the final state was, $e^{i(\theta_S+\theta_i)}|t+\tau_4\rangle_s|t+\tau_4\rangle_i + e^{i(2\omega*\tau_4)}|t+\tau_4\rangle_s|t+\tau_4\rangle_i$, where $\theta_S$, $\theta_i$, $\omega$ denote the MZI's phase in signal and idler's channel and the angular frequency of the pump light.

## 3. Experimental Setup

A tunable external cavity diode laser with 100 kHz linewidth was used to generate the pump light. The central wavelength of the laser was 1559 nm and the full width at half maximum (FWHM) of the coherence time was 4 μs. The output of the laser was amplified by two erbium-doped fiber amplifiers (EDFA) and then launched into a RPE PPLN waveguide to generate frequency-doubled pump light for the PDC process via second harmonic generation (SHG) as shown in Fig. 2. One 3.0-nm-wide tunable bandpass filters was inserted after each EDFA, respectively, to filter out the EDFA's spontaneous emission noise. Since these waveguide devices only accept TM-polarized light, an in-line fiber polarization controller was used to adjust the polarization of the input. The residual pump was attenuated by 180 dB using dichroic mirrors and pump filters. The second-harmonic (SH) wave was then launched into a second RPE PPLN waveguide to serve as the pump pulse for the PDC process. Waveguide-based PDC sources exhibit higher conversion efficiencies than bulk crystals and can be integrated with fiber-optic-based components.

Since RPE PPLN waveguides fabricated in z-cut substrates only support TM-polarized waves, separation of the near degenerate signal and idler waves via polarization de-multiplexing is not possible. To solve this problem, we took advantage of higher order mode interactions in combination with integrated mode multiplexers / demultiplexers via asymmetric Y-junctions [17, 18].

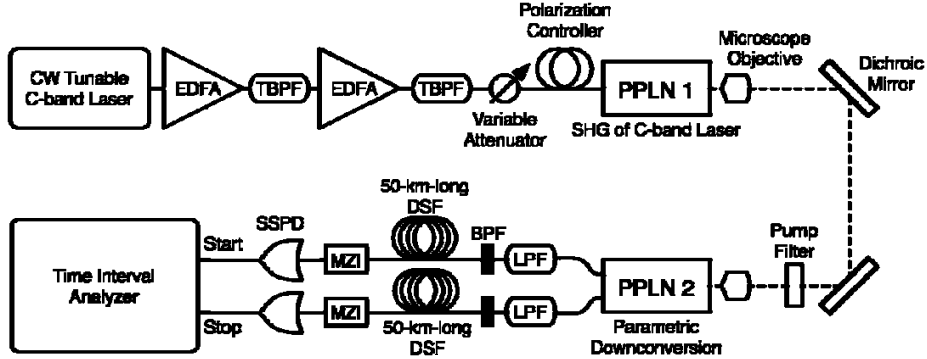

Fig. 2. Diagram of the experimental setup. TBPF: tunable band-pass filter. PPLN1: a RPE PPLN waveguide for second harmonic generation of the pump source. PPLN2: a fiber pigtailed asymmetric Y-junction RPE PPLN waveguide for parametric down-conversion. LPF: long-pass filter to remove the 780 nm pump light and other parasitics. BPF: 0.8-nm-wide bandpass filter. SSPD: superconducting single-photon detector. TIA: time interval analyzer. Solid lines represent optical fibers and dotted lines represent free-space propagation.

The generated entangled photon pairs and residual pump light were collected at the output of the RPE PPLN waveguide via a fiber pigtail. The spectral bandwidth of the entangled photon pairs was around 40 nm [17]. A long-pass filter was inserted into each output fiber of the second RPE PPLN chip to eliminate the residual pump light as well as other fluorescence. Two 0.8-nm-wide bandpass filters were used to reduce the bandwidth of the photon pairs, limiting the dispersion in the DSF. The band-pass filter also defined the photon pairs' time duration to be 4 ps FWHM. The photon pairs were then input into two spools of 50-km-long DSF, respectively. The center wavelength of the photon pairs was 1559 nm and the dispersion after 50 km broadened the photon pairs' pulse duration to 25 ps [19], which was smaller than the timing jitter of the SSPD.

The entangled photon pairs were analyzed by two 10-GHz unbalanced planar lightwave circuit (PLC) MZIs. The phase between the two arms of the MZIs could be controlled by adjusting the temperature of the interferometer using a Peltier element [20]. The time difference between the two arms was 100 ps, i. e. $\tau_4 = 100\,ps$, meeting the condition $\tau_2 < \tau_4 \ll \tau_1$. Therefore, one observes the two-photon-coincidence condition and not the single photon interference. Compared with previous entanglement distribution experiments, our MZIs had a significantly higher bandwidth, i.e. much smaller $\tau_4$, which could improve the entanglement flux.

However, a larger bandwidth needs a faster single-photon detector, i.e. $\tau_3 < \tau_4$. In most of the previous LDQC experiments, the detector's performance limited the distribution distance.

In our experiment, we used two SSPDs to detect the entangled photon pairs. The SSPDs used in this experiment consisted of a 100-nm-wide, 4-nm-thick NbN superconducting wire, which was coupled to a 9-μm core single-mode fiber [14]. The packaged detectors were housed in a closed-cycle cryogen-free refrigerator with an operating temperature of 3 K. The quantum efficiency and dark count rate of the SSPDs depended on an adjustable bias current. In the experiment, we set the bias current to reach a quantum efficiency of 0.7% and 2.1% for the signal and idler channel, respectively, and a dark count rate of around 100 Hz. These SSPDs had an inherently small Gaussian timing jitter of 65 ps (FWHM). In the experiment, we set our coincidence time window to 100 ps. Therefore, the dark count propability per time window was $10^{-8}$, which reduced the accidental coincidence rate caused by dark counts compared to previous experiments [7-9].

The detected signals were sent to a time interval analyzer (TIA) whose timing response was also faster than $\tau_4$ to measure the time coincidence histogram. The signal-photon's detection event were used as the start and the idler channel as the stop signal.

## 4. Experimental Results

In the experiment, we input 316 mW of pump at a wavelength of 1559 nm into the first PPLN waveguide from the EDFA and coupled 560 μW of the generated SH into the second PPLN chip.

We set the time window to 60 ps ($\tau_3$), which was equal to the timing jitter of the SSPD. Under this condition, we achieved 0.05 average photon pairs per time window. The total channel loss before the 100-km-long fiber was 20 dB, with 10 dB due to the PPLN's propagation, reflection, scattering and fiber pigtailing loss, and 10 dB loss from the filters and fiber U-benches. Each of the two PLC MZIs had a 5 dB insertion loss. The DSF's loss was 0.2 dB/km. Including all the channel loss terms, we obtained entangled photon pairs at a 2 Hz rate.

To qualify the entanglement after 100 km of fiber, we first set the PLC interferometer in the signal channel to 22.5°C and varied the temperature of the interferometer in the idler channel to obtain a coincidence interference-fringe pattern. To demonstrate entanglement, one interference pattern is not enough; at least one other pattern in a non-orthogonal basis is necessary. To observe this pattern, we set the signal interferometer to 24.5°C and observed the interference fringes shown in Fig. 4(a). The two curves with an average visibility of $(80.5 \pm 7)\%$, which is well beyond the visibility of 71% necessary for violation of the Bell inequality [21], demonstrate the achieved entanglement. The large uncertainty of the visibility resulted from system instabilities, for example, temperature fluctuations of the two PPLN chips, the polarization and timing fluctuations caused by the temperature change of the 100-km long fiber, etc. To eliminate the timing fluctuations, we increased the coincidence time window from 60 ps to 100 ps, even though this increased the accidental coincidence noise. Besides the timing fluctuations, the imperfection of the PLC MZIs, the dark counts of the SSPD and multi-photon-pair emissions will also introduce error to the visibility. In the experiment, it took several hours to take each curve in Fig. 4(a) due to the low flux of entangled photon pairs. The long measurement time accentuated polarization and timing fluctuations between the two measurements, and is also the reason for the different heights of the two curves in the figure.

To qualify the influence from the fiber, we also implemented a back-to-back visibility measurement without DSF as shown in Figure 3(b). Here, the average visibility was $(83.58 \pm 0.05)\%$. The imperfection of this visibility is due to accidental coincidence from the multi-photon-pair emission.

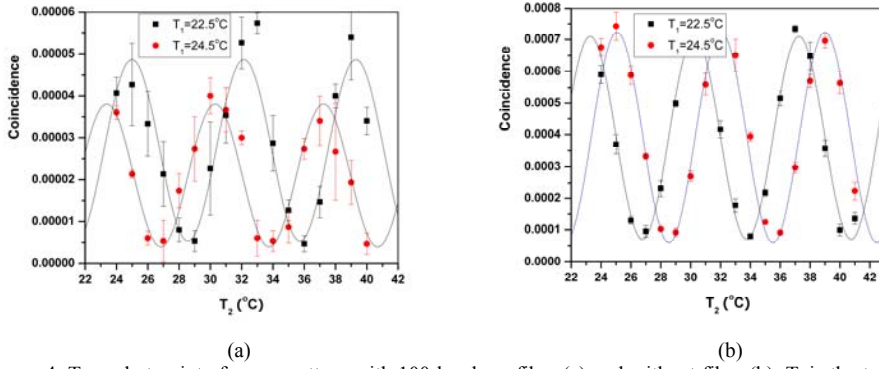

(a)            (b)

Figure 4. Two-photon interference pattern with 100-km-long fiber (a) and without fiber (b). $T_1$ is the temperature of the PLC MZI in the signal channel while $T_2$ is the temperature in the idler channel. The Y-axis represents the coincidence rate per signal photon with an average of 0.5 million signal photons.

## 5. Conclusion

We experimentally generated and distributed time-energy entanglement at telecom wavelengths with a RPE PPLN waveguide over a 100-km-long fiber with the help of SSPDs. The visibility of the achieved entanglement was $(80.5 \pm 7)\%$ without subtracting any noise when the entanglement flux was 2 Hz after 100 km.

## Acknowledgement

This research was supported by NICT, the U.S. Air Force Office of Scientific Research through contracts F49620-02-1-0240, the MURI center for photonic quantum information systems (ARO/ARDA program DAAD19-03-1-0199), the Disruptive Technology Office (DTO), SORST, and CREST programs, Japan Science and Technology Agency (JST) and the NIST quantum information science initiative. We acknowledge the support of Crystal Technology, Inc.